\documentclass[a4paper,11pt]{article}
\pdfoutput=1
\usepackage{jcappub}
\usepackage{longtable,tabu}
\usepackage{array}
\newcolumntype{P}[1]{>{\centering\arraybackslash}p{#1}}
\usepackage{graphicx}
\usepackage{epstopdf}
\usepackage{mathrsfs}
\usepackage{amssymb}
\usepackage{verbatim}
\usepackage{color}
\usepackage[dvipsnames]{xcolor}
\usepackage{multirow}
\usepackage{amsmath}
\usepackage{subfig}
\usepackage{slashed}
\usepackage{float}
\usepackage[normalem]{ulem}
\usepackage{sidecap}
\usepackage{mathtools}

\usepackage{diagbox}

\newcommand{\beq}{\begin{equation}}
\newcommand{\eeq}{\end{equation}}
\newcommand{\bea}{\begin{eqnarray}}
\newcommand{\eea}{\end{eqnarray}}

\newcommand{\zp}{\text{$Z^\prime$}}

\newcommand{\rh}{\text{\tiny{RH}}}
\newcommand{\Max}{\text{\tiny{MAX}}}

\newcommand{\const}{\text{\tiny{const}}}

\usepackage[parfill]{parskip}

\numberwithin{equation}{section}
\numberwithin{figure}{section}
\numberwithin{table}{section}

\title{\boldmath Dark matter seeping through dynamic gauge kinetic mixing}

\author[a]{Avik Banerjee,}
\author[a]{Gautam Bhattacharyya,}
\author[b,c]{Debtosh Chowdhury,}
\author[c]{Yann Mambrini}

\affiliation[a]{Saha Institute of Nuclear
    Physics, HBNI, 1/AF Bidhan Nagar, Kolkata 700064, India}
\affiliation[b]{Centre de Physique Th{\'e}orique, CNRS, {\'E}cole Polytechnique, IP Paris, 91128 Palaiseau, France}
\affiliation[c]{Laboratoire de Physique Th{\'e}orique (UMR8627), CNRS, Univ.~Paris-Sud, Universit{\'e} Paris-Saclay, 91405 Orsay, France}

\emailAdd{avik.banerjeesinp@saha.ac.in}
\emailAdd{gautam.bhattacharyya@saha.ac.in}
\emailAdd{debtosh.chowdhury@polytechnique.edu}
\emailAdd{yann.mambrini@th.u-psud.fr}

\abstract{
We show for the first time that the loop-driven kinetic mixing between visible and dark Abelian gauge bosons can facilitate dark matter production in the early Universe by creating a `dynamic' portal, which depends on the energy of the process. The required smallness of the strength of the portal interaction, suited for freeze-in, is justified by a suppression arising from the mass of a heavy vector-like fermion. The strong temperature sensitivity associated with the interaction is responsible for most of the dark matter production during the early stages of reheating.}

\arxivnumber{1905.11407}

\begin{document}
\maketitle
\flushbottom


\section{Introduction}
\label{intro}

More than 85 years ago Fritz Zwicky set a cat among the pigeons when he concluded in his seminal paper \cite{Zwicky:1933gu} that `dark matter is present in much greater amount than luminous matter' in the Coma cluster. Volumes of indirect confirmations such as combinations of the CMB measurements \cite{Aghanim:2018eyx} and astrophysical observations \cite{Riess:1998cb,Markevitch:2003at} although provide enough evidences for the existence of dark matter (DM) in the total energy budget of the Universe, the nature of the DM is yet to be understood. Due to its simplicity, strong predictability and naturalness, the Weakly Interacting Massive Particle (WIMP) paradigm has dominated the debate in dark matter searches and modeling during the last decades. From supersymmetric candidates to Kaluza-Klein excitations, there were plethora of motivations to justify that dark matter freezes out from the primordial plasma after a long stage of thermal equilibrium. 

The lack of DM detection in direct search experiments like XENON100 \cite{xenon100_collaboration_dark_2012_ok}, LUX \cite{akerib_results_2017_ok}, PandaX-II \cite{Cui:2017nnn} or more recently XENON1T \cite{Aprile:2018dbl}, however, drives us to look for alternative scenarios. Combined constraints from cosmology, direct searches and accelerator based experiments have already pushed the simplest extensions of the Standard Model ($Z$-portal \cite{ellis_statistical_2018,arcadi_z-portal_2015,kearney_$z$_2017,escudero_toward_2016}, Higgs-portal \cite{casas_reopening_2017,djouadi_implications_2012,djouadi_direct_2013,lebedev_vector_2012,mambrini_higgs_2011,Gross:2015cwa}, $Z'$-portal \cite{alves_dark_2014,lebedev_axial_2014,arcadi_invisible_2014,dudas_extra_2013,dudas_extra_2012,mambrini_zz_2011,Mambrini:2010dq} etc.) to unnatural corners of the parameter space  (see \cite{arcadi_waning_2018} for  recent reviews). This situation has led to the emergence of an alternative paradigm where the dark matter is conceived to be produced `in' the process of progressing towards thermal equilibrium, rather than being perceived as frozen `out' from the thermal bath. In order to avoid unacceptably large DM production resulting in over-closure of the universe, rather feeble couplings between the dark and the visible sectors are required. The Feebly Interacting Massive Particle (FIMP) scenario \cite{Hall:2009bx,Chu:2011be}, thus advocated is hardly a `miracle' unless the small couplings can be justified from an underlying dynamics. One such option is a mass-suppressed coupling, such as Planck scale suppressed couplings in supergravity as shown in  \cite{benakli_minimal_2017,dudas_case_2017,Dudas:2018npp,dudas_inflation_2017}, where the gravitino production is just sufficient to respect cosmological constraints in high-scale supersymmetric scenarios. In SO(10) unified theories, massive gauge bosons can play the role of heavy mediators yielding also small couplings \cite{mambrini_dark_2015,Mambrini:2013iaa,Mambrini:2016dca}. Similar suppressions also arise in massive spin-2 theories \cite{bernal_spin-2_2018,garny_theory_2017}, string theory inspired moduli portal scenarios \cite{Chowdhury:2018tzw} and in scenarios containing Chern-Simons type couplings \cite{bhattacharyya_freezing-dark_2018}. A notable feature in all these constructions is a sharp temperature dependence of the DM relic density -- beyond the conventional reheating temperature ($T_{\rh}$) -- up to some maximum temperature ($T_{\Max}$) accessible during the reheating process \cite{garcia_enhancement_2017,garcia_pre-thermalization_2018}. As an aside, we mention here that DM production through freeze-in can also happen directly from the inflaton decay \cite{Kaneta:2019zgw}.

Another possibility, that we show for the first time in this paper, is freeze-in DM production through radiatively generated gauge kinetic mixing. Portals of kinetic mixing with constant strengths have often been used in the literature in the context of various UV complete scenarios \cite{Holdom:1985ag,Dienes:1996zr, Abel:2008ai, Goodsell:2009xc} to motivate DM production \cite{Feldman:2006wd,Kang:2010mh,Chu:2013jja,Mambrini:2011dw}. On the contrary, in our case, the portal between a dark $\rm U(1)^\prime$ and hypercharge $\rm U(1)_Y$, generated by loops of some heavy vector-like fermion exhibits a strong temperature dependence (hence, `dynamic'), and can effectively produce dark matter in sufficient amount in the early stages of the reheating. The extreme smallness of the coupling is guaranteed in this case by the suppression arising from the heaviness of the loop fermion together with the loop factor. 

The paper is organized as follows. In Section~\ref{model}, we describe our model and calculate the radiatively generated dynamic gauge kinetic mixing. We then compute and analyze the DM relic abundance in Section~\ref{freezein} before concluding in Section~\ref{concl}.


\section{Dynamic kinetic mixing portal}
\label{model}

\subsection{The model}

We consider the following scenario to illustrate the emergence of dynamic gauge kinetic mixing between two Abelian sectors. We assume the presence of a vector mediator $\zp$ coupled to a fermionic DM $\chi$ while keeping the Standard Model sector neutral with respect to it. This $\zp$ can arise from gauging a $\rm U(1)^\prime$ and may receive a mass ($M_{\zp}$) by St\"uckelberg or some dark Higgs mechanism. The Lagrangian of the dark sector containing a massive $\zp$ is then given by 
\begin{equation}
\label{lag_dark}
\mathcal{L}_{\textrm{dark}}=-\dfrac{1}{ 4}\zp^{\mu\nu}\zp_{\mu\nu}+\dfrac{1}{2}M_\zp^2\zp^{\mu}\zp_{\mu}+\bar{\chi}(i\slashed D-m_\chi)\chi\, ,
\end{equation}
where $\slashed D=\slashed\partial+ i g_{D}q_\chi\slashed Z^\prime$ 
and $\zp_{\mu \nu}= \partial_\mu \zp_\nu-\partial_\nu \zp_\mu$ is the field strength of $\zp$. Following the principle of gauge invariance, one can write a tree level kinetic mixing term between the dark $\rm U(1)^\prime$ and the hypercharge $\rm U(1)_Y$, given by
\begin{equation}
\label{lag_mix}
\mathcal{L}_{\textrm{mix}}=-\dfrac{\delta}{2}B^{\mu\nu}\zp_{\mu\nu},
\end{equation}
\noindent
$B_\mu$ being the gauge field associated with the Standard Model hypercharge. The literature is rich in studies where $\delta$ is a free parameter, generally small\footnote{This smallness corresponds to a tuning arising from some UV dynamics. In particular, a UV realization of vanishing tree level kinetic mixing has been envisaged in the literature \cite{Dienes:1996zr} if either of the two $\rm U(1)$ factors transcends from a non-Abelian group. Radiative effects, however, will give rise to finite logarithmic corrections to the kinetic mixing \cite{Holdom:1985ag}.} to avoid overproduction of dark matter in freeze-out or freeze-in scenarios, while in the mean time respecting direct detection constraints. In what follows, we will assume that the two Abelian sectors dominantly communicate through some hybrid mediators. Similar, if not identical, situations arise in GUT models which accommodate heavy fermions. As a consequence, we neglect the tree level (contact) mixing in our framework to study the effect of the radiatively generated kinetic mixing. Here in passing we mention that a possible realistic UV setup leading to tiny contact mixing term may arise from a {\it clockwork} mechanism, as already been studied extensively in the literature \cite{Giudice:2016yja,Lee:2017fin,Gherghetta:2019coi}. In Appendix \ref{clockwork} we present the clockwork mechanism for generating negligibly small kinetic mixing parameter.  

In our scenario, the hybrid mediators are a set of heavy fermions $F_j$, which are vector-like under  both $\rm U(1)^\prime$ and $\rm U(1)_Y$. The Lagrangian in this sector may be written as
\begin{equation}
\label{lag_hyb}
\mathcal{L}_{\textrm{hybrid}}=\sum_{j}^{N_F}\bar{F_j}(i\slashed\partial-m_{j}-g^\prime Q^\prime_j\slashed B-g_DQ_{Dj}\slashed \zp)F_j\, ,
\end{equation}
\noindent
where $N_F$ is the number of hybrid fermions and we assume that $m_j\gg M_\zp$. For simplicity and without lack of generalities, we consider a minimal setup where $N_F=1$, $m_j=m_F$, $Q^\prime_j=Q^\prime$ and $Q_{Dj}=Q_D$. We now proceed to compute the gauge kinetic mixing generated by this fermion at energy scales below $m_F$. 

\subsection{Emergence of dynamic gauge kinetic mixing}

\begin{figure}[t!]
	\centering
	\includegraphics[trim = 0mm 0mm 0mm 7mm,clip,width=0.4\textwidth]{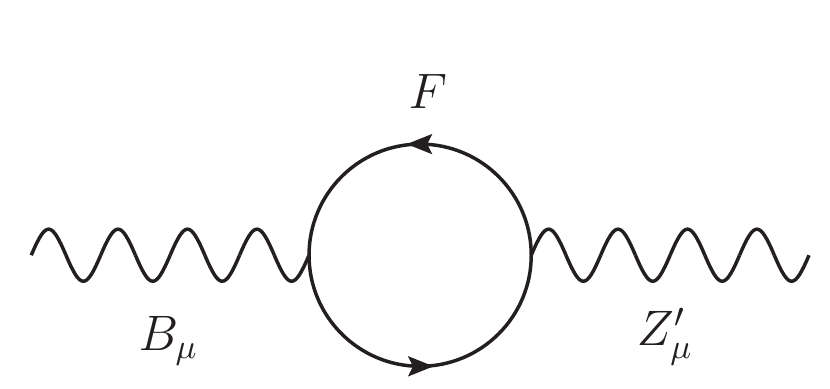}
	\caption{\small One loop graph for kinetic mixing.}
	\label{Fig:kinmix}
\end{figure}
Once the heavy hybrid fermion is integrated out, an effective kinetic mixing is radiatively generated (see Fig.~\ref{Fig:kinmix}) for processes occurring at energies below $m_F$. Note that, the corresponding one loop mixed vacuum polarization diagram shown in Fig.~\ref{Fig:kinmix} contains a logarithmically divergent piece. Since the mixing term corresponds to a marginal gauge invariant operator, even if we have neglected the tree level mixing as mentioned previously, a dimension-4 counterterm exists in the absence of any forbidding symmetry to take care of the divergence. The one loop contribution from Fig.~\ref{Fig:kinmix} has the structure (see Appendix~\ref{one loop} for the complete expression)
\begin{eqnarray}
i\Pi^{\mu\nu}_{\zp B}(p^2)=i\Pi_{\zp B}(p^2)\left(p^2\eta^{\mu\nu}-p^\mu p^\nu\right)\, ,
\end{eqnarray}
where $\Pi_{\zp B}$, calculated using the Dimensional Regularization scheme in the limit $p^2\ll m_F^2$, with $\mu$ as the renormalization scale, is given by 
\begin{align}
\label{mix01}
\Pi_{\zp B}(p^2)\simeq-\dfrac{(g^\prime Q^\prime)(g_DQ_D)}{12\pi^2}\left[{1\over \hat{\epsilon}}+\log\left({\mu^2 \over m_F^2}\right)\right.
+\left.{p^2\over 5m_F^2}+\mathcal{O}\left({p^4\over m_F^4}\right)\right]\, .
\end{align}
The renormalized kinetic mixing for $p^2\ll m_F^2$ is then
\begin{eqnarray}
\delta_{\textrm{ren}}(p^2)=\Pi_{\zp B}(p^2)-\delta_{\textrm{CT}}\, ,
\end{eqnarray}
where $\delta_{\textrm{CT}}$ denotes the counterterm. We recall that $g^\prime$ and $g_D$ will have usual logarithmic running triggered by the standard and dark degrees of freedom, respectively. We nevertheless fix them to constant values, as the effect of their running is numerically insignificant for the purpose of our analysis. The natural renormalization prescription we employ for the determination of the counterterm is that at large distance ($p^2\to 0$) the mixing vanishes to keep the quantum electrodynamics totally uncontaminated. This implies that
\begin{equation}
\delta_{\textrm{ren}}(0)=\Pi_{\zp B}(0)-\delta_{\textrm{CT}}=0\, .
\end{equation}
It immediately follows that
\begin{align}
\label{mix02}
\delta_{\textrm{ren}}(p^2)&=\Pi_{\zp B}(p^2)-\Pi_{\zp B}(0)\simeq-\dfrac{(g^\prime Q^\prime)(g_DQ_D)}{60\pi^2}{p^2\over m_F^2}+\mathcal{O}\left({p^4\over m_F^4}\right)\, .
\end{align}
The above expression is reminiscent of the origin of Lamb shift in quantum electrodynamics. Effectively, the counterterm absorbs the logarithmic correction in addition to the divergent piece. On the other hand, in momentum independent renormalization schemes ({\it e.g.} $\overline{{\rm MS}}$ scheme) one sets $\mu=m_F$ to implement the decoupling of heavy hybrid particles in the loop \cite{Manohar:1996cq}, leading to the same final result as given in Eq.~\eqref{mix02}. 
Thus the effective kinetic mixing below the hybrid fermion mass scale is of the order $\mathcal{O}(p^2/m_F^2)$ reduced by a loop factor\footnote{In particular, if either of the $\rm U(1)$ factors has a non-Abelian parentage in the the UV realization as indicated in \cite{Holdom:1985ag,Dienes:1996zr}, cancellation of one loop divergence is ensured without the presence of any counterterm, in addition to the vanishing of tree level kientic mixing. However, in this specific scenario the momentum dependent mixing will continue to remain sub-leading in comparison to the logarithmic contribution. Therefore, we do not appeal to this usual UV realization of embedding one of the $\rm U(1)$ factors into a non-Abelian group to promote the relevance of the momentum dependent portal. Instead, we alluded to the presence of a clockwork mechanism at the UV responsible for generating negligible contact mixing.}. Additionally, due to the explicit momentum dependence involved, the strength of the mixing depends on the scale and dynamics of the process under consideration. These two attributes make such dynamic mixing a worthy portal for freezing-in DM.

Note that, at low energy the loop contribution can be envisaged through the following dimension-6 operator,
\begin{equation}
\label{op_dim_6}
\mathcal{O}^{(6)}_{\zp B}=\dfrac{1}{\Lambda_{\textrm{eff}}^2}B_{\mu\nu}\Box\zp^{\mu\nu},\quad
\textrm{with}\quad
\dfrac{1}{\Lambda_\textrm{eff}^2}=\dfrac{(g^\prime Q^\prime)(g_DQ_D)}{60\pi^2}\dfrac{1}{m_F^2}.
\end{equation}
    

\section{Freezing-in dark matter}
\label{freezein}

To calculate the evolution of dark matter number density ($n_\chi$)
we need the Boltzmann equation:
\beq \label{Eq:dndt}
\frac{\text{d} n_\chi}{\text{d}t} = -3 H(t) n_\chi + R(T)\, ,
\eeq
where $R(T)$ denotes the dark matter production rate and $H(T)$ is the usual Hubble expansion rate. In our scenario, two main production channels are the following: \textit{(i)} $f\bar{f}\to \chi\bar{\chi}$ and \textit{(ii)} $H^\dagger H\to \chi\bar{\chi}$, where $f$ and $H$ denote the Standard Model fermions and Higgs doublet, respectively. 

We emphasize that the contribution of the inflaton field ($\phi$) to the total energy density can dominate over that of radiation if $M_\zp$ is close to reheating temperature ($T_\rh$). In that case the dark matter relic density is calculated by solving Eq.~\eqref{Eq:dndt} along with the following two equations for the inflaton field and the radiation\footnote{Notice that, we do not consider direct production of dark matter from inflaton decay in the present scenario \cite{Kaneta:2019zgw}.} \cite{giudice_largest_2000,garcia_enhancement_2017}:
\begin{align}\label{Eq:setboltzmann}
\frac{\text{d}\rho_\gamma}{\text{d}t} &\approx -4H\,\rho_\gamma+\Gamma_\phi\,\rho_\phi\, , \notag \\ 
\frac{\text{d}\rho_\phi}{\text{d}t} &= -3H\,\rho_\phi-\Gamma_\phi\,\rho_\phi\, ,
\end{align}
where we have neglected dark matter interaction with radiation in the evolution of  radiation energy density. The solution of these coupled differential equations can be well approximated analytically in the limiting cases of inflaton and radiation domination. For radiation dominated era the standard expression involving the Hubble rate is given by
\begin{equation}
\frac{\text{d}}{\text{d}t} = - H(T) T
\frac{\text{d}}{\text{d}T} \quad \textrm{with}~~ H(T) =
\sqrt{\frac{g_e}{90}} \pi \frac{T^2}{M_P},
\end{equation}
while the same for the inflaton dominated era is given by \cite{mazumdar_quantifying_2014,bhattacharyya_freezing-dark_2018}
\begin{equation}
\frac{\text{d}}{\text{d}t} = - \frac{3}{8} H(T) T
\frac{\text{d}}{\text{d}T} \quad \textrm{with}~~ H(T) =
\sqrt{\frac{5 g_\Max^2}{72 g_\rh}} \pi \frac{T^4}{T_\rh^2 M_P}.
\end{equation}
Here, $g_\rh$ and $g_\Max$ represent the relativistic degrees of freedom at $T_\rh$ and at the maximal temperature ($T_\Max$) reached during the reheating process, respectively, and $M_P=2.8\times 10^{18}$ GeV is the reduced Planck mass. We will assume the energetic and entropic relativistic degrees of freedom, $g_e$ and $g_s$, are equal to $106.75$. Using the above equations, the dark matter relic density $\Omega h^2 \equiv m_\chi n_\chi/\rho_{c}$ (where $\rho_{c}$ is the critical density today) can be calculated by splitting it into two parts \textit{viz.} a radiation dominated and an inflaton dominated contributions, as \cite{Chowdhury:2018tzw}
\begin{equation}
\begin{split}\label{relicsplit}
\Omega h^2 \cong \Omega h^2_{RD} + \Omega h^2_{ID} \sim 4 \times 10^{24}~m_\chi \left( \int_{T_0}^{T_\rh} dT \frac{R(T)}{T^6} \right.\left. +\, 1.07 ~T_\rh^7 \int_{T_\rh}^{T_\Max} dT \frac{R(T)}{T^{13}} \right),   
\end{split}
\end{equation}
where $T_0$ is the present temperature. It turns out that the production of the dark matter will have dominant contribution from the inflaton dominated era if the temperature dependence of the rate follows as $R(T)\propto T^n$ with $n\geq 12$. In the following analysis, we will assume $T_\Max=100\, T_\rh$ for the purpose of illustration.

\subsection{Production rate}

We present below the generic structure of the dark matter production rates obtained in our model for three distinct ranges of $M_\zp$, assuming $m_\chi\ll T$, as
\begin{equation}\label{rate01}
R (T) = \frac{\mathcal{C}}{\left(4\pi\right)^4} \times \left\{ \begin{array}{lc}
\dfrac{T^{8}}{m_F^4}\, , ~~~~~~~~~~~~~~~~~~~~~~~ (M_\zp \ll T) \\ \noalign{\medskip} \dfrac{M_{\zp}^8}{m_F^4}\dfrac{T}{\Gamma_{\zp}}K_1\left(\dfrac{M_{\zp}}{T}\right)\, , ~~ (M_\zp \sim T) \\ \noalign{\medskip} 
\dfrac{T^{12}}{m_F^4M_{\zp}^4}\, , ~~~~~~~~~~~~~~~~~~ (M_\zp \gg T) 
\end{array} \right.
\end{equation}
\begin{figure}[t!]
\centering
\includegraphics[width=0.6\textwidth]{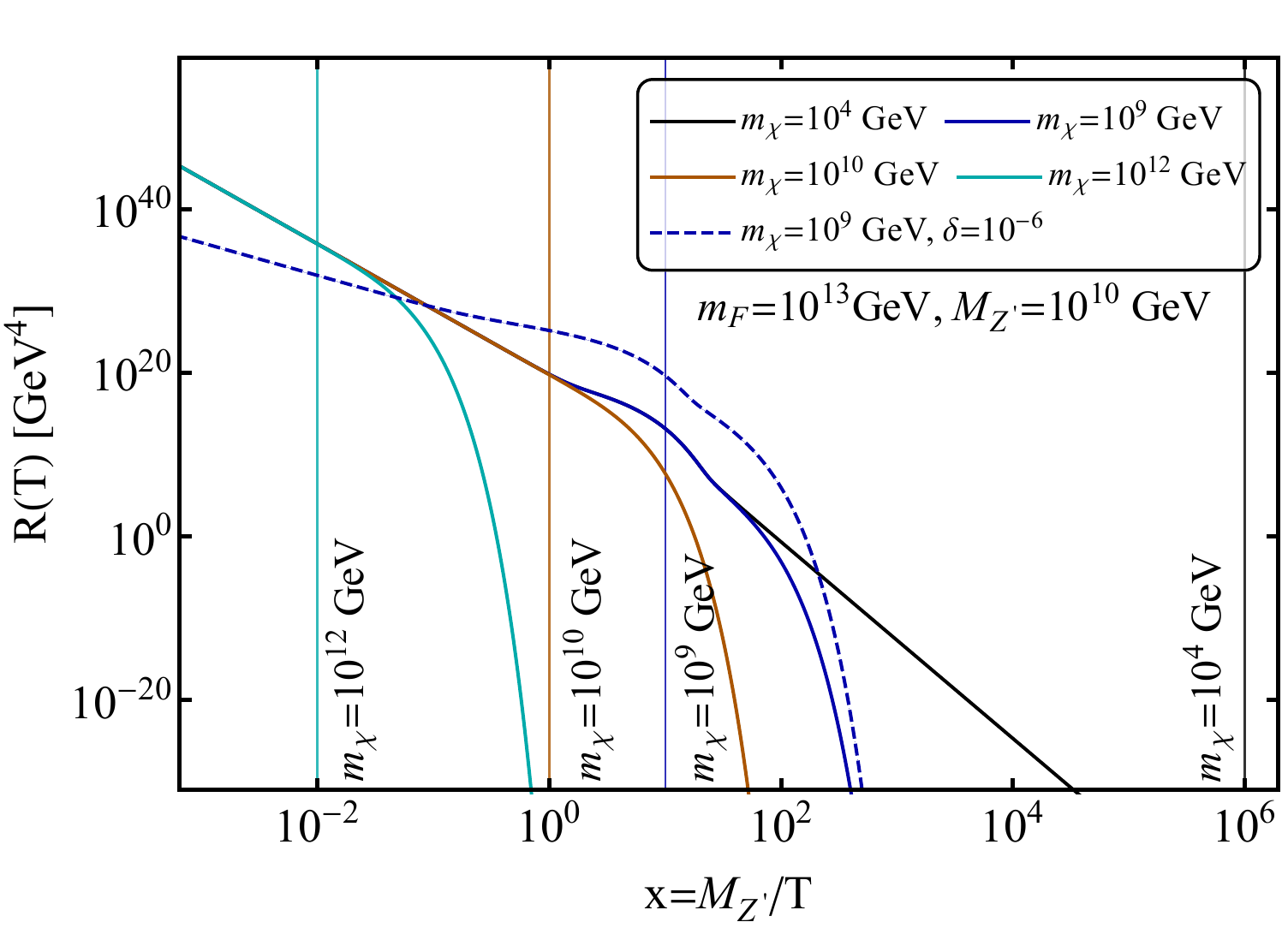}
\caption{\small DM production rate for both dynamic (solid curves) and constant (dashed curve) kinetic mixing portals.}
\label{Fig:rate}
\end{figure}
where we take the decay width $\Gamma_\zp\ll M_\zp$ (see Appendix~\ref{const} for the detailed expressions of $R(T)$, $\Gamma_{\zp}$ and the numerical coefficients $\mathcal{C}$). We perform a full numerical computation of the rate using the CUBA package \cite{hahn_cuba_2005}. For a set of benchmark parameters, the DM production rate as a function of the variable $x\equiv M_\zp/T$ is displayed in Fig.~\ref{Fig:rate} (solid curves). For the ease of illustration we have set $g_D^2q_\chi Q^{\prime}Q_D=1$ (see Eqs.~\eqref{lag_dark} and \eqref{lag_hyb}), $m_F = 10^{13}$ GeV, and $M_{\zp} = 10^{10}$ GeV.  From left to right in the solid curves, $m_\chi=10^{12}$, $10^{10}$, $10^{9}$, and $10^{4}$ GeV (cyan, brown, blue, and black), respectively. From the expressions of the approximate rates in Eq.\eqref{rate01}, we can intuitively follow the different regimes of DM production shown in Fig.~\ref{Fig:rate}. The production rate has a pronounced temperature dependence and in general falls as the universe cools down. In the small $x \ll 1$ (large $T$) regime, the bath temperature is much higher than the mediator mass, and hence the rate is governed by the light mediator approximation ($M_{\zp} \ll T$). In the large $x \gg 1$ (small $T$) regime, sufficient temperature is not available in the bath to produce $\zp$ on-shell, indicating the region dictated by the heavy mediator approximation ($M_{\zp} \gg T$). However, if the bath temperature is around the $\zp$ mass ($x \sim 1$), dark matter is produced through the on-shell $\zp$ decay leading to $s$-channel resonance enhancement. Thus, the $\zp$-pole effects are observed around $x\sim 1$ and the production rate is governed by the narrow width approximation ($M_{\zp} \sim T$). Furthermore, once the temperature falls below $m_\chi$, the production rate drops exponentially due to the well-known Boltzmann suppression ($\propto e^{-m_\chi/T}$). Colored vertical lines mark $T=m_\chi$ for the four different values of the dark matter masses. For $m_\chi=10^{12}$ GeV and $10^{10}$ GeV, Boltzmann suppression predates the $\zp$ pole. For these cases, resonance enhancement around $x \sim 1$ is absent in the production rate. 

We also compare in Fig.~\ref{Fig:rate} the DM production rates as found in our model with that found using a tree level constant kinetic mixing\footnote{Strictly speaking, $\delta$, as defined in Eq.~\eqref{lag_mix}, does not remain a constant but  runs logarithmically being proportional to itself. For the purpose of comparison, we treat $\delta$ as a constant, as the numerical effect of its running on the DM production is negligible.} portal (dashed blue curve) for $m_\chi = 10^9$ GeV and kinetic mixing parameter $\delta = 10^{-6}$. In the latter case, the temperature dependence of the production rates for different $M_\zp$ are given by
\begin{equation}\label{rate02}
R^\const (T) = \mathcal{C}^\const \times \left\{ \begin{array}{lc}
\delta^2 T^4\, , ~~~~~~~~~~~~~~~~~~~~~~~ (M_\zp \ll T) \\ \noalign{\medskip} \delta^2M_{\zp}^4\dfrac{T}{\Gamma_{\zp}}K_1\left(\dfrac{M_{\zp}}{T}\right)\, , ~ (M_\zp \sim T) \\ \noalign{\medskip} 
\delta^2 \dfrac{T^{8}}{M_{\zp}^4}\, ,~~~~~~~~~~~~~~~~~~~~~ (M_\zp \gg T) 
\end{array} \right.
\end{equation} 
where the coefficients $\mathcal{C}^{\text{\tiny{const}}}$ are given in Appendix~\ref{const}. The comparison shows that in case of constant kinetic mixing, as the bath temperature decreases, the production rate falls at a slower pace than for dynamic mixing. This aspect can be accounted by noting the relative suppressions between Eqs.~\eqref{rate01} and \eqref{rate02}. Thus, while for the dynamic portal the DM will be produced mostly at early times leading to a UV freeze-in, the production will take place for a prolonged duration in the constant mixing scenario depending on the strength of the mixing parameter.

\subsection{Relic abundance}

\begin{figure}[t!]
\centering
\includegraphics[width=0.6\textwidth]{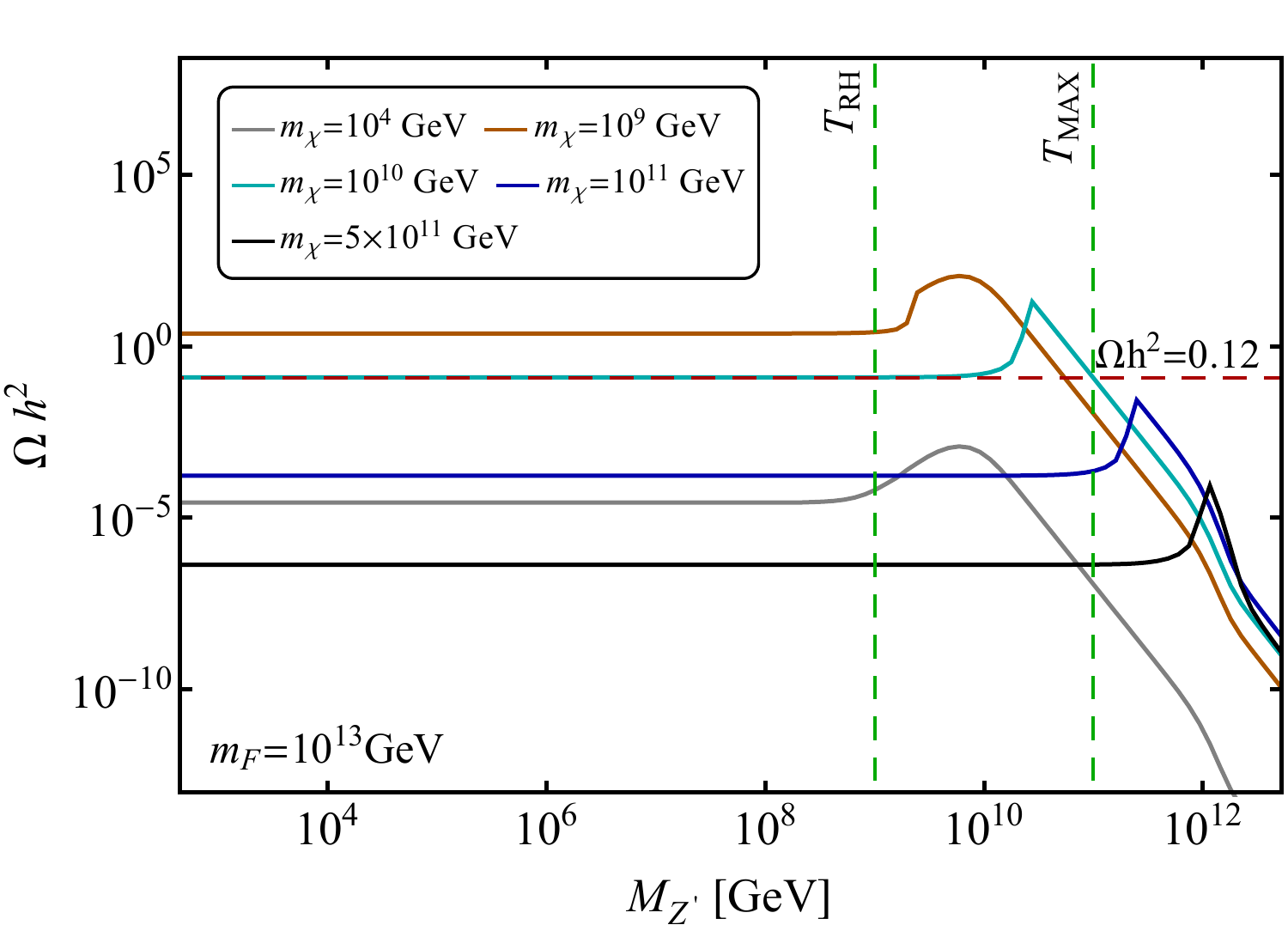}
\caption{\small Dependence of DM relic abundance on $\zp$ mass for dynamic portal.}
\label{Fig:omega-mzp}
\end{figure}

We now calculate the DM relic abundance in our model, and examine the consequences of matching the relic density to the observed value $\Omega h^2\sim 0.12$. In Fig.~\ref{Fig:omega-mzp}, we exhibit the dependence of the relic density on $M_\zp$ for different values of $m_\chi$ (colored solid lines). In the light mediator regime ($M_\zp\ll T_\rh$), $\Omega h^2$ is insensitive to $M_\zp$ as the relic abundance saturates at a much higher temperature. In the $T_\rh\lesssim M_\zp\lesssim T_\Max$ region the relic density increases due to $s$-channel resonance when $M_\zp\simeq2m_\chi$. When we consider heavier $\zp$ its on-shell production from the bath gets suppressed causing a fall in the relic abundance. Once $M_\zp\gg T_\Max$ the density falls more sharply. To understand the dependence of the relic density on the DM mass, we recall that $\Omega h^2\propto m_\chi n_\chi$. For relatively smaller values of $m_\chi$ the abundance grows with increasing $m_\chi$ (gray and brown curves), while we witness a fall in $\Omega h^2$ once $m_\chi$ goes above $T_\rh$ (cyan, blue and black curves) via a severe phase space suppression in $n_\chi$.

\begin{figure}[t!]
\centering
\includegraphics[width=0.6\textwidth]{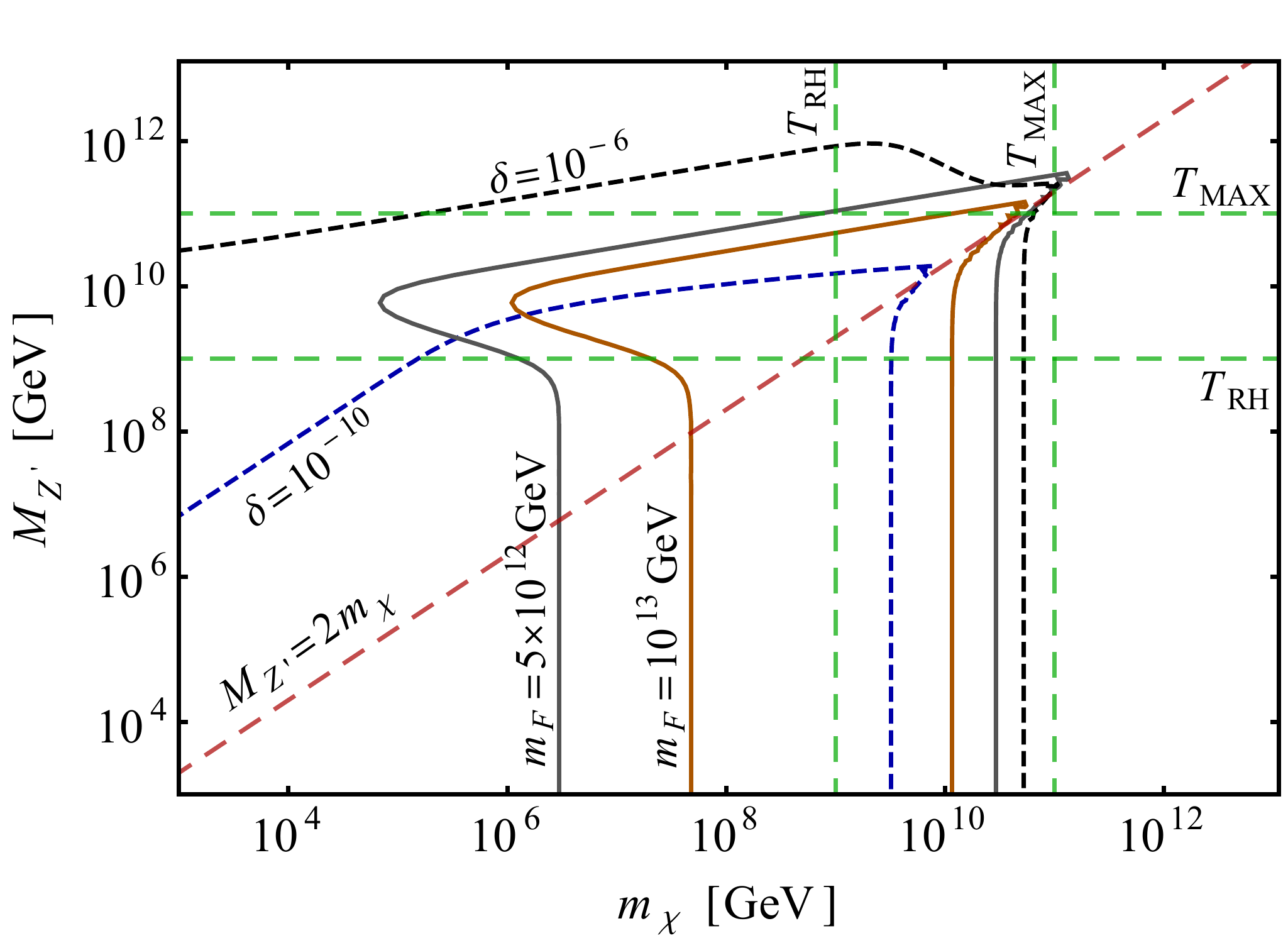}
\caption{\small Contours of $\Omega h^2 = 0.12$ for both dynamic (brown and gray solid curves) and constant (black and blue dashed curves) mixing portals.}
\label{Fig:mdm-mzp}
\end{figure}

In Fig.~\ref{Fig:mdm-mzp}, we present the contours of $\Omega h^2=0.12$ in the $M_\zp-m_\chi$ plane for both dynamic and constant kinetic mixing portals. We first discuss the dynamic kinetic mixing results as obtained in our model for two representative choices of $m_F=5\times 10^{12}$ GeV (gray) and $10^{13}$ GeV (brown), respectively. Each choice of $m_F$ corresponds to a contour, on which $m_\chi n_\chi$ is constant, implying that lighter (heavier) DM needs to be produced in large (small) number. More specifically, the right (left)-hand branch of the contour is associated with less (more) DM production. For low $M_\zp$ ($\ll T_\rh$) the contour is insensitive to $M_\zp$ as explained in the context of Fig.~\ref{Fig:omega-mzp}. When $M_\zp\sim T_\rh$, excess DM production due to resonance is counterbalanced as the left-handed branch of the contour (which was so long vertical) turns towards smaller $m_\chi$. The contour cannot continue indefinitely towards increasingly smaller $m_\chi$ as $n_\chi$ needs to be appropriately compensated by arranging a lighter mediator (\textit{i.e.}\, small $M_\zp$), which in turn weakens the dynamic portal ($\propto M_\zp^2/m_F^2$). This explains the upper left edge of the contour. The  contour then turns right towards larger $m_\chi$ requiring  monotonically increasing $M_\zp$ to keep $m_\chi n_\chi$ to a constant value. Finally beyond certain values of $m_\chi$ and $M_\zp$, the DM production is insufficient to reproduce the observed relic, justifying the upper right edge of the contour. We also observe that the contour for $m_F=10^{13}$ GeV is contained within that of $m_F=5\times10^{12}$ GeV, which can be explained by simply noting that larger (smaller) $m_F$ implies weaker (stronger) kinetic mixing ($\propto 1/m_F^2$). At this point we make a quantitative estimate of the required smallness of the contact term in comparison to the $p^2-$dependent term for different regions of parameter space in Fig.~\ref{Fig:mdm-mzp}, to justify the viability of the above discussion. Comparing Eqs.~\eqref{rate01} and \eqref{rate02} we obtain the condition to render the effects of the contact term negligible as
\begin{equation}\label{relic01}
\delta \ll \frac{1}{16\pi^2} \frac{T^2}{m_F^2}.
\end{equation} 
Since the relic density gets saturated at or above $T\sim m_\chi$, for $M_{Z^\prime}\ll T_\rh$, $m_\chi\sim 10^6$ GeV and $m_F\sim10^{12}$ GeV we estimate $\delta\ll 10^{-14}$ is required to be neglected safely. On the other hand, for $M_{Z^\prime}\geq T_\rh$ the condition relaxes to a great extent to give $\delta\ll 10^{-8}$.

For comparison, we also ran our analysis with constant kinetic mixing contours for $\delta=10^{-6}$ (black dashed), and $10^{-10}$ (blue dashed). The primary difference with the dynamic portal case is the absence of additional powers of temperature endowed in the dynamics. For a given $\delta$, the vertical line is absent in the left-hand side as a large $M_\zp$ is required to tame the DM over production. Larger $\delta$ obviously requires heavier $\zp$ to reproduce the relic density. For $\delta=10^{-6}$, when $m_\chi$ crosses $T_\rh$, Boltzmann suppression shows up in the form of a dip. This happens because in the constant mixing case the DM production occurs almost entirely in the radiation dominated era, in contrast with the dynamic mixing scenario where additional powers of $T$ is responsible for DM production even in the inflaton dominated period ($T_\rh<T<T_\Max$). For $\delta=10^{-10}$, once $M_\zp$ crosses $T_\rh$ the slope of the contour changes to adjust $m_\chi n_\chi=$ constant. 
     
\section{Conclusions and outlook}
\label{concl}

The most noteworthy observation in this paper is the identification of a scale-dependent portal for freezing-in DM production. The portal is created through one loop gauge kinetic mixing between a dark $\rm U(1)^\prime$ and hypercharge $\rm U(1)_Y$ by integrating out a very heavy vector-like fermion. The requirement of preserving quantum electrodynamics at large distances entails the strength of this mixing strongly dependent on the energy of the process involved. This novel route, not conceived previously, allows the dark matter to be produced through freeze-in mechanism mostly during the very early stage of reheating. We have demonstrated how it differs from freeze-in DM production through constant kinetic mixing. It is worth stressing that in the absence of tree level kinetic mixing, that can be attributed to some tuning, the mixing arising in our model provides the required smallness of the portal interaction, side by side with an enhanced temperature dependence leading to a UV freeze-in. Needless to add, though `freeze-in' was primarily motivated to justify the continued absence of evidence in DM direct searches, it is time to put serious thoughts on any possible, however far-fetched, tests of such scenarios. For instance, possible future detection of gravitational waves, generated if the $\rm U(1)^\prime$ breaking is associated with first order phase transition \cite{Jaeckel:2016jlh,Dev:2016feu,Hasegawa:2019amx}, may point towards a $\zp$ mass range far beyond the reach of any future colliders, thus shedding some light on the DM portal.     
An interesting corollary would be to investigate whether the concept of this dynamic kinetic mixing can be employed in a `freeze-out' scenario, \textit{albeit} with a different range of parameters \cite{freeze_out_kin_mix}.

\acknowledgments
We profusely thank Emilian Dudas for enlightening discussions. DC would like to thank Thomas Hambye, Michel Tytgat and James D. Wells for useful discussions. This research has been supported by the (Indo-French) CEFIPRA/IFCPAR Project No.~5404-2. Support from CNRS LIA-THEP and the INFRE-HEPNET of CEFIPRA/IFCPAR is also acknowledged. AB acknowledges financial support from the Department of Atomic Energy, Government of India. GB acknowledges support of the J.C. Bose National Fellowship from the Department of Science and Technology, Government of India (SERB Grant No. SB/S2/JCB-062/2016). DC would like to thank Universit{\'e} Libre de Bruxelles theory group for their hospitality during the final stages of the work. This work was also supported by the France-US PICS no.~06482, PICS MicroDark. This project has received funding/support from the European Union’s Horizon 2020 research and innovation programme under the Marie Sk\l{}odowska-Curie: RISE InvisiblesPlus (grant agreement No.~690575) and the ITN Elusives (grant agreement No.~674896).


\appendix

\section{Tiny kinetic mixing {\it \`a la} clockwork mechanism}
\label{clockwork}

Clockwork setup consists of $N+1$ gauged $\rm U(1)$ symmetries spontaneously broken to a single $\rm U(1)$ at very high scale ($f\gg m_F$) by vacuum expectation values of $N$ scalar link fields \cite{Giudice:2016yja}. Each of these scalar fields are charged under two neighbouring sites with charges $(1,-q)$. The corresponding Lagrangian involving the gauge fields below the scale $f$ is given by
\begin{equation}
\mathcal{L}=-\sum_{k=0}^{N}\frac{1}{4}F^k_{\mu\nu}F^{k\mu\nu}+\sum_{k=0}^{N-1}\frac{g_c^2f^2}{2}\left(A^k_\mu-qA^{k+1}_\mu\right)^2.
\end{equation}
After diagonalization to the mass basis, $N$ massive gauge bosons ($\tilde{A}^k_\mu$) with masses of the order of $g_cf\gg m_F$ are produced, keeping one gauge boson ($Z^\prime_\mu$) light corresponding to the unbroken $\rm U(1)$. We identify the latter with our $\rm U(1)^\prime$. Mass of the $Z^\prime$ can be generated at much lower scales independent of the clockwork mechanism, as mentioned earlier. The gauge fields at the $N^{\rm th}$ site ($A^N_\mu$) and at the zeroth site ($A^0_\mu$) can be expressed in terms of the mass basis as 
\begin{equation}
A^N_\mu= \frac{N_0}{q^N}Z^\prime_\mu+\sum_{k=1}^{N}a_{Nk}\tilde{A}^k_\mu,\qquad
A^0_\mu= N_0Z^\prime_\mu+\sum_{k=1}^{N}a_{0k}\tilde{A}^k_\mu~,
\end{equation}
where $N_0$ is an $\mathcal{O}(1)$ constant and $a_{jk}$ denotes the elements of diagonalizing matrix with $\mathcal{O}(1)$ values. Clearly, if $B_\mu$ has a dimension-4 kinetic mixing with $A^N_\mu$ only, by virtue of the clockwork mechanism, the $Z^\prime$ will have geometrically suppressed mixing at the tree level \cite{Giudice:2016yja,Lee:2017fin,Gherghetta:2019coi}, given by
\begin{equation}
\delta\sim\frac{\mathcal{O}(1)}{q^N}.
\end{equation} 
With large number of sites, this framework provides a working example where the tree level mixing can be neglected in comparison to the radiative contributions coming from different sources, thereby justifying our choice mentioned in the text. On the other hand, the other heavy clockwork modes ($\tilde{A}^k_\mu$) despite having large mixing with $B_\mu$ has negligible contribution to dark matter phenomenology as long as their masses are much larger than $m_F$. Therefore, we can safely integrate out these heavy modes keeping only $Z^\prime$ as relevant dynamic gauge field coming from the clockwork framework. Unlike the hypercharge, the dark matter and the hybrid mediators are assumed to couple to the clockwork setup only at the zeroth site ({\it i.e.} with $A^0_\mu$). As a result $Z^\prime$ will see
the DM and the hybrid mediator with $\mathcal{O}(1)$ interaction strength. 

\section{Calculation of one loop diagram }
\label{one loop}

The one loop vacuum polarization diagram, shown in Fig.~\ref{Fig:kinmix}, is calculated using the Dimensional Regularization scheme ($d=4-2\epsilon$) as follows:
\begin{small}
\begin{align}
i\Pi_{\zp B}^{\mu\nu}(p^2)=-\int \dfrac{d^{d}k}{ (2\pi)^{d}}\dfrac{\textrm{Tr}\left[(g^\prime Q^\prime\gamma^\mu)(\slashed k+m_F)(g_DQ_D\gamma^\nu)(\slashed k-\slashed p+m_F)\right]}{ \left[k^2-m_F^2\right]\left[(k-p)^2-m_F^2\right]}= i\Pi_{\zp B}(p^2)\left(p^2\eta^{\mu\nu}-p^\mu p^\nu\right).
\end{align}
\end{small}
The full analytic expression for $\Pi_{\zp B}$ is given in terms of $r=p^2/4m_F^2$ as 
\begin{small}
\begin{align}
\nonumber
\Pi_{\zp B}(p^2) &= -\dfrac{(g^\prime Q^\prime)(g_DQ_D)}{ 12\pi^2}\left[\dfrac{1}{\hat{\epsilon}}+\log\left(\dfrac{\mu^2}{ m_F^2}\right)+\dfrac{5}{3}+\dfrac{1}{r}+\sqrt{1-\dfrac{1}{r}}\left(1+\dfrac{1}{2r}\right)\log\left(1-2r+2\sqrt{r(r-1)}\right)\right] \, , \\
\label{oneloop01}
&\mathrel{\stackrel{r\ll 1}{\simeq}}-\dfrac{(g^\prime Q^\prime)(g_DQ_D)}{ 12\pi^2}
\left[\frac{1}{\hat{\epsilon}}+ \log \left( \frac{ \mu^2}{m_F^2} \right) + \frac{5}{3} + \frac{1}{r} - 2\sqrt{\frac{1}{r}-1}\left(1 + \frac{1}{2r}\right) \sin^{-1} \left(\sqrt{r}\right)
\right]\, ,
\end{align}
\end{small}
where
\[
\frac{1}{\hat{\epsilon}} \equiv \frac{1}{\epsilon} - \gamma_{E} + \log 4\pi \, ,
\]
and $\gamma_E\simeq 0.577$ is the Euler-Mascheroni constant. Evidently as $r\to 0$, Eq.~\eqref{oneloop01} reduces to Eq.~\eqref{mix01}.


\section{Expressions for $R(T)$ and $\Gamma_\zp$}
\label{const}

\begin{table*}[t!]
\begin{tabular}{|c||c|c|}
\hline \rule[-2ex]{0pt}{5.5ex} $\mathcal{C}$ & $f\bar{f}\to \chi\bar{\chi}$ &  $H^\dagger H\to \chi\bar{\chi}$\\ 
\hline \hline \rule[-2.5ex]{0pt}{6.5ex} $M_\zp\ll T$ & $\dfrac{1568g^{\prime 4}\beta^2}{675\pi^5}$ &  $\dfrac{16g^{\prime 4}\beta^2}{225\pi^5}$ \\ 
\hline \rule[-2.5ex]{0pt}{6.5ex} $M_\zp\sim T$ & $\dfrac{49g^{\prime 4}\beta^2}{16200\pi^4}$ &  $\dfrac{g^{\prime 4}\beta^2}{10800\pi^4}$ \\ 
\hline \rule[-2.5ex]{0pt}{6.5ex} $M_\zp\gg T$ & $\dfrac{401408g^{\prime 4}\beta^2}{45\pi^5}$ & $\dfrac{4096g^{\prime 4}\beta^2}{15\pi^5}$ \\ 
\hline 
\end{tabular} 
\hspace{1cm}
\begin{tabular}{|c||c|c|}
\hline \rule[-2ex]{0pt}{5.5ex} $\mathcal{C}^{\text{\tiny{const}}}$ & $f\bar{f}\to \chi\bar{\chi}$ &  $H^\dagger H\to \chi\bar{\chi}$\\ 
\hline \hline \rule[-2.5ex]{0pt}{6.5ex} $M_\zp\ll T$ & $\dfrac{49g^{\prime 2}\beta^{\prime 2}}{288\pi^5}$ &  $\dfrac{g^{\prime 2}\beta^{\prime 2}}{192\pi^5}$ \\ 
\hline \rule[-2.5ex]{0pt}{6.5ex} $M_\zp\sim T$ & $\dfrac{49g^{\prime 2}\beta^{\prime 2}}{1152\pi^4}$ &  $\dfrac{g^{\prime 2}\beta^{\prime 2}}{768\pi^4}$ \\ 
\hline \rule[-2.5ex]{0pt}{6.5ex} $M_\zp\gg T$ & $\dfrac{98g^{\prime 2}\beta^{\prime 2}}{3\pi^5}$ & $\dfrac{g^{\prime 2}\beta^{\prime 2}}{\pi^5}$ \\
\hline 
\end{tabular} 
\caption{Expressions for the coefficients $\mathcal{C}$ and $\mathcal{C}^{\text{\tiny{const}}}$, where $\beta\equiv g_D^2q_\chi Q^{\prime}Q_D$ and $\beta^\prime\equiv g_Dq_\chi$.}
\label{tab_const}
\end{table*}
The expression for the rate of DM production, defined in Eq.~\eqref{Eq:dndt}, is given by
\begin{equation}
R(T)=\alpha \left(g^\prime g_D q_\chi\right)^2 T \int_{4m_\chi^2}^{\infty}ds \sqrt{s-4m_\chi^2}\, K_1\left(\frac{\sqrt{s}}{T}\right)\delta^2_\textrm{ren}(s)\frac{s(s+2m_\chi^2)}{(s-M_\zp^2)^2+M_\zp^2\Gamma_\zp^2}\, ,
\end{equation}
where $K_1(x)$ denotes modified Bessel function of the second kind and $\delta_{\textrm{ren}}(s)$ can be read off from Eq.~\eqref{mix02} for dynamic mixing. The coefficient $\alpha$ for the production channels \textit{(i)} $f\bar{f}\to \chi\bar{\chi}$ and \textit{(ii)} $H^\dagger H\to \chi\bar{\chi}$ are, respectively given by
\begin{equation}
\alpha_{f\bar{f}\to \chi\bar{\chi}}=\frac{1}{96\pi^5}\sum_{f}\left(a_f^2+v_f^2\right), \quad \alpha_{H^\dagger H\to\chi\bar{\chi}}=\frac{1}{768\pi^5}\, ,
\end{equation}
where $a_f$ and $v_f$ are the vector and axial-vector couplings of the visible fermions with $B_\mu$. In case of quarks in the initial state, an additional factor in $\alpha$, due to the number of colors ($N_c=3$) should be taken into account.
Numerical constants $\mathcal{C}$ and $\mathcal{C}^{\text{\tiny{const}}}$ appearing in Eqs.~\eqref{rate01} and \eqref{rate02} for the two production channels are displayed in Table~\ref{tab_const}.

We assume that the decay width of $\zp$ to the Standard Model particles are small compared to that to the dark matter, due to the smallness of kinetic mixing. The expression for the decay width of $\zp$ to a pair of dark matter particles is given by
\begin{equation}
\Gamma_{\zp}=\dfrac{g_D^2q_\chi^2}{12\pi}M_\zp\left(1+\dfrac{2m_\chi^2}{M_\zp^2}\right)\sqrt{1-\dfrac{4m_\chi^2}{M_\zp^2}}\, .
\end{equation}


\bibliographystyle{JHEP}
\bibliography{main}

\providecommand{\href}[2]{#2}\begingroup\raggedright\begin{thebibliography}{10}

\bibitem{Zwicky:1933gu}
F.~Zwicky, \emph{{Die Rotverschiebung von extragalaktischen Nebeln}},
  \href{http://dx.doi.org/10.1007/s10714-008-0707-4}{\emph{Helv. Phys. Acta}
  {\bf 6} (1933) 110--127}.

\bibitem{Aghanim:2018eyx}
{\scshape Planck} collaboration, N.~Aghanim et~al., \emph{{Planck 2018 results.
  VI. Cosmological parameters}},  \href{https://arxiv.org/abs/1807.06209}{{\tt
  1807.06209}}.

\bibitem{Riess:1998cb}
{\scshape Supernova Search Team} collaboration, A.~G. Riess et~al.,
  \emph{{Observational evidence from supernovae for an accelerating universe
  and a cosmological constant}},
  \href{http://dx.doi.org/10.1086/300499}{\emph{Astron. J.} {\bf 116} (1998)
  1009--1038}, [\href{https://arxiv.org/abs/astro-ph/9805201}{{\tt
  astro-ph/9805201}}].

\bibitem{Markevitch:2003at}
M.~Markevitch, A.~H. Gonzalez, D.~Clowe, A.~Vikhlinin, L.~David, W.~Forman
  et~al., \emph{{Direct constraints on the dark matter self-interaction
  cross-section from the merging galaxy cluster 1E0657-56}},
  \href{http://dx.doi.org/10.1086/383178}{\emph{Astrophys. J.} {\bf 606} (2004)
  819--824}, [\href{https://arxiv.org/abs/astro-ph/0309303}{{\tt
  astro-ph/0309303}}].

\bibitem{xenon100_collaboration_dark_2012_ok}
{\scshape XENON100} collaboration, E.~Aprile et~al., \emph{{Dark Matter Results
  from 225 Live Days of XENON100 Data}},
  \href{http://dx.doi.org/10.1103/PhysRevLett.109.181301}{\emph{Phys. Rev.
  Lett.} {\bf 109} (2012) 181301}, [\href{https://arxiv.org/abs/1207.5988}{{\tt
  1207.5988}}].

\bibitem{akerib_results_2017_ok}
{\scshape LUX} collaboration, D.~S. Akerib et~al., \emph{{Results from a search
  for dark matter in the complete LUX exposure}},
  \href{http://dx.doi.org/10.1103/PhysRevLett.118.021303}{\emph{Phys. Rev.
  Lett.} {\bf 118} (2017) 021303},
  [\href{https://arxiv.org/abs/1608.07648}{{\tt 1608.07648}}].

\bibitem{Cui:2017nnn}
{\scshape PandaX-II} collaboration, X.~Cui et~al., \emph{{Dark Matter Results
  From 54-Ton-Day Exposure of PandaX-II Experiment}},
  \href{http://dx.doi.org/10.1103/PhysRevLett.119.181302}{\emph{Phys. Rev.
  Lett.} {\bf 119} (2017) 181302},
  [\href{https://arxiv.org/abs/1708.06917}{{\tt 1708.06917}}].

\bibitem{Aprile:2018dbl}
{\scshape XENON} collaboration, E.~Aprile et~al., \emph{{Dark Matter Search
  Results from a One Ton-Year Exposure of XENON1T}},
  \href{http://dx.doi.org/10.1103/PhysRevLett.121.111302}{\emph{Phys. Rev.
  Lett.} {\bf 121} (2018) 111302},
  [\href{https://arxiv.org/abs/1805.12562}{{\tt 1805.12562}}].

\bibitem{ellis_statistical_2018}
J.~Ellis, A.~Fowlie, L.~Marzola and M.~Raidal, \emph{{Statistical Analyses of
  Higgs- and Z-Portal Dark Matter Models}},
  \href{http://dx.doi.org/10.1103/PhysRevD.97.115014}{\emph{Phys. Rev.} {\bf
  D97} (2018) 115014}, [\href{https://arxiv.org/abs/1711.09912}{{\tt
  1711.09912}}].

\bibitem{arcadi_z-portal_2015}
G.~Arcadi, Y.~Mambrini and F.~Richard, \emph{{Z-portal dark matter}},
  \href{http://dx.doi.org/10.1088/1475-7516/2015/03/018}{\emph{JCAP} {\bf 1503}
  (2015) 018}, [\href{https://arxiv.org/abs/1411.2985}{{\tt 1411.2985}}].

\bibitem{kearney_$z$_2017}
J.~Kearney, N.~Orlofsky and A.~Pierce, \emph{{$Z$ boson mediated dark matter
  beyond the effective theory}},
  \href{http://dx.doi.org/10.1103/PhysRevD.95.035020}{\emph{Phys. Rev.} {\bf
  D95} (2017) 035020}, [\href{https://arxiv.org/abs/1611.05048}{{\tt
  1611.05048}}].

\bibitem{escudero_toward_2016}
M.~Escudero, A.~Berlin, D.~Hooper and M.-X. Lin, \emph{{Toward (Finally!)
  Ruling Out Z and Higgs Mediated Dark Matter Models}},
  \href{http://dx.doi.org/10.1088/1475-7516/2016/12/029}{\emph{JCAP} {\bf 1612}
  (2016) 029}, [\href{https://arxiv.org/abs/1609.09079}{{\tt 1609.09079}}].

\bibitem{casas_reopening_2017}
J.~A. Casas, D.~G. Cerde{\~n}o, J.~M. Moreno and J.~Quilis, \emph{{Reopening
  the Higgs portal for single scalar dark matter}},
  \href{http://dx.doi.org/10.1007/JHEP05(2017)036}{\emph{JHEP} {\bf 05} (2017)
  036}, [\href{https://arxiv.org/abs/1701.08134}{{\tt 1701.08134}}].

\bibitem{djouadi_implications_2012}
A.~Djouadi, O.~Lebedev, Y.~Mambrini and J.~Quevillon, \emph{{Implications of
  LHC searches for Higgs--portal dark matter}},
  \href{http://dx.doi.org/10.1016/j.physletb.2012.01.062}{\emph{Phys. Lett.}
  {\bf B709} (2012) 65--69}, [\href{https://arxiv.org/abs/1112.3299}{{\tt
  1112.3299}}].

\bibitem{djouadi_direct_2013}
A.~Djouadi, A.~Falkowski, Y.~Mambrini and J.~Quevillon, \emph{{Direct Detection
  of Higgs-Portal Dark Matter at the LHC}},
  \href{http://dx.doi.org/10.1140/epjc/s10052-013-2455-1}{\emph{Eur. Phys. J.}
  {\bf C73} (2013) 2455}, [\href{https://arxiv.org/abs/1205.3169}{{\tt
  1205.3169}}].

\bibitem{lebedev_vector_2012}
O.~Lebedev, H.~M. Lee and Y.~Mambrini, \emph{{Vector Higgs-portal dark matter
  and the invisible Higgs}},
  \href{http://dx.doi.org/10.1016/j.physletb.2012.01.029}{\emph{Phys. Lett.}
  {\bf B707} (2012) 570--576}, [\href{https://arxiv.org/abs/1111.4482}{{\tt
  1111.4482}}].

\bibitem{mambrini_higgs_2011}
Y.~Mambrini, \emph{{Higgs searches and singlet scalar dark matter: Combined
  constraints from XENON 100 and the LHC}},
  \href{http://dx.doi.org/10.1103/PhysRevD.84.115017}{\emph{Phys. Rev.} {\bf
  D84} (2011) 115017}, [\href{https://arxiv.org/abs/1108.0671}{{\tt
  1108.0671}}].

\bibitem{Gross:2015cwa}
C.~Gross, O.~Lebedev and Y.~Mambrini, \emph{{Non-Abelian gauge fields as dark
  matter}}, \href{http://dx.doi.org/10.1007/JHEP08(2015)158}{\emph{JHEP} {\bf
  08} (2015) 158}, [\href{https://arxiv.org/abs/1505.07480}{{\tt 1505.07480}}].

\bibitem{alves_dark_2014}
A.~Alves, S.~Profumo and F.~S. Queiroz, \emph{{The dark $Z^{'}$ portal: direct,
  indirect and collider searches}},
  \href{http://dx.doi.org/10.1007/JHEP04(2014)063}{\emph{JHEP} {\bf 04} (2014)
  063}, [\href{https://arxiv.org/abs/1312.5281}{{\tt 1312.5281}}].

\bibitem{lebedev_axial_2014}
O.~Lebedev and Y.~Mambrini, \emph{{Axial dark matter: The case for an invisible
  $Z′$}}, \href{http://dx.doi.org/10.1016/j.physletb.2014.05.025}{\emph{Phys.
  Lett.} {\bf B734} (2014) 350--353},
  [\href{https://arxiv.org/abs/1403.4837}{{\tt 1403.4837}}].

\bibitem{arcadi_invisible_2014}
G.~Arcadi, Y.~Mambrini, M.~H.~G. Tytgat and B.~Zaldivar, \emph{{Invisible
  $Z^\prime$ and dark matter: LHC vs LUX constraints}},
  \href{http://dx.doi.org/10.1007/JHEP03(2014)134}{\emph{JHEP} {\bf 03} (2014)
  134}, [\href{https://arxiv.org/abs/1401.0221}{{\tt 1401.0221}}].

\bibitem{dudas_extra_2013}
E.~Dudas, L.~Heurtier, Y.~Mambrini and B.~Zaldivar, \emph{{Extra U(1),
  effective operators, anomalies and dark matter}},
  \href{http://dx.doi.org/10.1007/JHEP11(2013)083}{\emph{JHEP} {\bf 11} (2013)
  083}, [\href{https://arxiv.org/abs/1307.0005}{{\tt 1307.0005}}].

\bibitem{dudas_extra_2012}
E.~Dudas, Y.~Mambrini, S.~Pokorski and A.~Romagnoni, \emph{{Extra U(1) as
  natural source of a monochromatic gamma ray line}},
  \href{http://dx.doi.org/10.1007/JHEP10(2012)123}{\emph{JHEP} {\bf 10} (2012)
  123}, [\href{https://arxiv.org/abs/1205.1520}{{\tt 1205.1520}}].

\bibitem{mambrini_zz_2011}
Y.~Mambrini, \emph{{The ZZ' kinetic mixing in the light of the recent direct
  and indirect dark matter searches}},
  \href{http://dx.doi.org/10.1088/1475-7516/2011/07/009}{\emph{JCAP} {\bf 1107}
  (2011) 009}, [\href{https://arxiv.org/abs/1104.4799}{{\tt 1104.4799}}].

\bibitem{Mambrini:2010dq}
Y.~Mambrini, \emph{{The Kinetic dark-mixing in the light of CoGENT and
  XENON100}},
  \href{http://dx.doi.org/10.1088/1475-7516/2010/09/022}{\emph{JCAP} {\bf 1009}
  (2010) 022}, [\href{https://arxiv.org/abs/1006.3318}{{\tt 1006.3318}}].

\bibitem{arcadi_waning_2018}
G.~Arcadi, M.~Dutra, P.~Ghosh, M.~Lindner, Y.~Mambrini, M.~Pierre et~al.,
  \emph{{The waning of the WIMP? A review of models, searches, and
  constraints}},
  \href{http://dx.doi.org/10.1140/epjc/s10052-018-5662-y}{\emph{Eur. Phys. J.}
  {\bf C78} (2018) 203}, [\href{https://arxiv.org/abs/1703.07364}{{\tt
  1703.07364}}].

\bibitem{Hall:2009bx}
L.~J. Hall, K.~Jedamzik, J.~March-Russell and S.~M. West, \emph{{Freeze-In
  Production of FIMP Dark Matter}},
  \href{http://dx.doi.org/10.1007/JHEP03(2010)080}{\emph{JHEP} {\bf 03} (2010)
  080}, [\href{https://arxiv.org/abs/0911.1120}{{\tt 0911.1120}}].

\bibitem{Chu:2011be}
X.~Chu, T.~Hambye and M.~H.~G. Tytgat, \emph{{The Four Basic Ways of Creating
  Dark Matter Through a Portal}},
  \href{http://dx.doi.org/10.1088/1475-7516/2012/05/034}{\emph{JCAP} {\bf 1205}
  (2012) 034}, [\href{https://arxiv.org/abs/1112.0493}{{\tt 1112.0493}}].

\bibitem{benakli_minimal_2017}
K.~Benakli, Y.~Chen, E.~Dudas and Y.~Mambrini, \emph{{Minimal model of
  gravitino dark matter}},
  \href{http://dx.doi.org/10.1103/PhysRevD.95.095002}{\emph{Phys. Rev.} {\bf
  D95} (2017) 095002}, [\href{https://arxiv.org/abs/1701.06574}{{\tt
  1701.06574}}].

\bibitem{dudas_case_2017}
E.~Dudas, Y.~Mambrini and K.~Olive, \emph{{Case for an EeV Gravitino}},
  \href{http://dx.doi.org/10.1103/PhysRevLett.119.051801}{\emph{Phys. Rev.
  Lett.} {\bf 119} (2017) 051801},
  [\href{https://arxiv.org/abs/1704.03008}{{\tt 1704.03008}}].

\bibitem{Dudas:2018npp}
E.~Dudas, T.~Gherghetta, K.~Kaneta, Y.~Mambrini and K.~A. Olive,
  \emph{{Gravitino decay in high scale supersymmetry with R -parity
  violation}}, \href{http://dx.doi.org/10.1103/PhysRevD.98.015030}{\emph{Phys.
  Rev.} {\bf D98} (2018) 015030}, [\href{https://arxiv.org/abs/1805.07342}{{\tt
  1805.07342}}].

\bibitem{dudas_inflation_2017}
E.~Dudas, T.~Gherghetta, Y.~Mambrini and K.~A. Olive, \emph{{Inflation and
  High-Scale Supersymmetry with an EeV Gravitino}},
  \href{http://dx.doi.org/10.1103/PhysRevD.96.115032}{\emph{Phys. Rev.} {\bf
  D96} (2017) 115032}, [\href{https://arxiv.org/abs/1710.07341}{{\tt
  1710.07341}}].

\bibitem{mambrini_dark_2015}
Y.~Mambrini, N.~Nagata, K.~A. Olive, J.~Quevillon and J.~Zheng, \emph{{Dark
  matter and gauge coupling unification in nonsupersymmetric SO(10) grand
  unified models}},
  \href{http://dx.doi.org/10.1103/PhysRevD.91.095010}{\emph{Phys. Rev.} {\bf
  D91} (2015) 095010}, [\href{https://arxiv.org/abs/1502.06929}{{\tt
  1502.06929}}].

\bibitem{Mambrini:2013iaa}
Y.~Mambrini, K.~A. Olive, J.~Quevillon and B.~Zaldivar, \emph{{Gauge Coupling
  Unification and Nonequilibrium Thermal Dark Matter}},
  \href{http://dx.doi.org/10.1103/PhysRevLett.110.241306}{\emph{Phys. Rev.
  Lett.} {\bf 110} (2013) 241306}, [\href{https://arxiv.org/abs/1302.4438}{{\tt
  1302.4438}}].

\bibitem{Mambrini:2016dca}
Y.~Mambrini, N.~Nagata, K.~A. Olive and J.~Zheng, \emph{{Vacuum Stability and
  Radiative Electroweak Symmetry Breaking in an SO(10) Dark Matter Model}},
  \href{http://dx.doi.org/10.1103/PhysRevD.93.111703}{\emph{Phys. Rev.} {\bf
  D93} (2016) 111703}, [\href{https://arxiv.org/abs/1602.05583}{{\tt
  1602.05583}}].

\bibitem{bernal_spin-2_2018}
N.~Bernal, M.~Dutra, Y.~Mambrini, K.~Olive, M.~Peloso and M.~Pierre,
  \emph{{Spin-2 Portal Dark Matter}},
  \href{http://dx.doi.org/10.1103/PhysRevD.97.115020}{\emph{Phys. Rev.} {\bf
  D97} (2018) 115020}, [\href{https://arxiv.org/abs/1803.01866}{{\tt
  1803.01866}}].

\bibitem{garny_theory_2017}
M.~Garny, A.~Palessandro, M.~Sandora and M.~S. Sloth, \emph{{Theory and
  Phenomenology of Planckian Interacting Massive Particles as Dark Matter}},
  \href{http://dx.doi.org/10.1088/1475-7516/2018/02/027}{\emph{JCAP} {\bf 1802}
  (2018) 027}, [\href{https://arxiv.org/abs/1709.09688}{{\tt 1709.09688}}].

\bibitem{Chowdhury:2018tzw}
D.~Chowdhury, E.~Dudas, M.~Dutra and Y.~Mambrini, \emph{{Moduli Portal Dark
  Matter}}, \href{http://dx.doi.org/10.1103/PhysRevD.99.095028}{\emph{Phys.
  Rev.} {\bf D99} (2019) 095028}, [\href{https://arxiv.org/abs/1811.01947}{{\tt
  1811.01947}}].

\bibitem{bhattacharyya_freezing-dark_2018}
G.~Bhattacharyya, M.~Dutra, Y.~Mambrini and M.~Pierre, \emph{{Freezing-in dark
  matter through a heavy invisible Z′}},
  \href{http://dx.doi.org/10.1103/PhysRevD.98.035038}{\emph{Phys. Rev.} {\bf
  D98} (2018) 035038}, [\href{https://arxiv.org/abs/1806.00016}{{\tt
  1806.00016}}].

\bibitem{garcia_enhancement_2017}
M.~A.~G. Garcia, Y.~Mambrini, K.~A. Olive and M.~Peloso, \emph{{Enhancement of
  the Dark Matter Abundance Before Reheating: Applications to Gravitino Dark
  Matter}}, \href{http://dx.doi.org/10.1103/PhysRevD.96.103510}{\emph{Phys.
  Rev.} {\bf D96} (2017) 103510}, [\href{https://arxiv.org/abs/1709.01549}{{\tt
  1709.01549}}].

\bibitem{garcia_pre-thermalization_2018}
M.~A.~G. Garcia and M.~A. Amin, \emph{{Prethermalization production of dark
  matter}}, \href{http://dx.doi.org/10.1103/PhysRevD.98.103504}{\emph{Phys.
  Rev.} {\bf D98} (2018) 103504}, [\href{https://arxiv.org/abs/1806.01865}{{\tt
  1806.01865}}].

\bibitem{Kaneta:2019zgw}
K.~Kaneta, Y.~Mambrini and K.~A. Olive, \emph{{Radiative production of
  nonthermal dark matter}},
  \href{http://dx.doi.org/10.1103/PhysRevD.99.063508}{\emph{Phys. Rev.} {\bf
  D99} (2019) 063508}, [\href{https://arxiv.org/abs/1901.04449}{{\tt
  1901.04449}}].

\bibitem{Holdom:1985ag}
B.~Holdom, \emph{{Two U(1)'s and Epsilon Charge Shifts}},
  \href{http://dx.doi.org/10.1016/0370-2693(86)91377-8}{\emph{Phys. Lett.} {\bf
  166B} (1986) 196--198}.

\bibitem{Dienes:1996zr}
K.~R. Dienes, C.~F. Kolda and J.~March-Russell, \emph{{Kinetic mixing and the
  supersymmetric gauge hierarchy}},
  \href{http://dx.doi.org/10.1016/S0550-3213(97)80028-4,
  10.1016/S0550-3213(97)00173-9}{\emph{Nucl. Phys.} {\bf B492} (1997)
  104--118}, [\href{https://arxiv.org/abs/hep-ph/9610479}{{\tt
  hep-ph/9610479}}].

\bibitem{Abel:2008ai}
S.~A. Abel, M.~D. Goodsell, J.~Jaeckel, V.~V. Khoze and A.~Ringwald,
  \emph{{Kinetic Mixing of the Photon with Hidden U(1)s in String
  Phenomenology}},
  \href{http://dx.doi.org/10.1088/1126-6708/2008/07/124}{\emph{JHEP} {\bf 07}
  (2008) 124}, [\href{https://arxiv.org/abs/0803.1449}{{\tt 0803.1449}}].

\bibitem{Goodsell:2009xc}
M.~Goodsell, J.~Jaeckel, J.~Redondo and A.~Ringwald, \emph{{Naturally Light
  Hidden Photons in LARGE Volume String Compactifications}},
  \href{http://dx.doi.org/10.1088/1126-6708/2009/11/027}{\emph{JHEP} {\bf 11}
  (2009) 027}, [\href{https://arxiv.org/abs/0909.0515}{{\tt 0909.0515}}].

\bibitem{Feldman:2006wd}
D.~Feldman, B.~Kors and P.~Nath, \emph{{Extra-weakly Interacting Dark Matter}},
  \href{http://dx.doi.org/10.1103/PhysRevD.75.023503}{\emph{Phys. Rev.} {\bf
  D75} (2007) 023503}, [\href{https://arxiv.org/abs/hep-ph/0610133}{{\tt
  hep-ph/0610133}}].

\bibitem{Kang:2010mh}
Z.~Kang, T.~Li, T.~Liu, C.~Tong and J.~M. Yang, \emph{{Light Dark Matter from
  the $U(1)_X$ Sector in the NMSSM with Gauge Mediation}},
  \href{http://dx.doi.org/10.1088/1475-7516/2011/01/028}{\emph{JCAP} {\bf 1101}
  (2011) 028}, [\href{https://arxiv.org/abs/1008.5243}{{\tt 1008.5243}}].

\bibitem{Chu:2013jja}
X.~Chu, Y.~Mambrini, J.~Quevillon and B.~Zaldivar, \emph{{Thermal and
  non-thermal production of dark matter via Z'-portal(s)}},
  \href{http://dx.doi.org/10.1088/1475-7516/2014/01/034}{\emph{JCAP} {\bf 1401}
  (2014) 034}, [\href{https://arxiv.org/abs/1306.4677}{{\tt 1306.4677}}].

\bibitem{Mambrini:2011dw}
Y.~Mambrini, \emph{{The ZZ' kinetic mixing in the light of the recent direct
  and indirect dark matter searches}},
  \href{http://dx.doi.org/10.1088/1475-7516/2011/07/009}{\emph{JCAP} {\bf 1107}
  (2011) 009}, [\href{https://arxiv.org/abs/1104.4799}{{\tt 1104.4799}}].

\bibitem{Giudice:2016yja}
G.~F. Giudice and M.~McCullough, \emph{{A Clockwork Theory}},
  \href{http://dx.doi.org/10.1007/JHEP02(2017)036}{\emph{JHEP} {\bf 02} (2017)
  036}, [\href{https://arxiv.org/abs/1610.07962}{{\tt 1610.07962}}].

\bibitem{Lee:2017fin}
H.~M. Lee, \emph{{Gauged $U(1)$ clockwork theory}},
  \href{http://dx.doi.org/10.1016/j.physletb.2018.01.010}{\emph{Phys. Lett.}
  {\bf B778} (2018) 79--87}, [\href{https://arxiv.org/abs/1708.03564}{{\tt
  1708.03564}}].

\bibitem{Gherghetta:2019coi}
T.~Gherghetta, J.~Kersten, K.~Olive and M.~Pospelov, \emph{{The Price of Tiny
  Kinetic Mixing}},  \href{https://arxiv.org/abs/1909.00696}{{\tt 1909.00696}}.

\bibitem{Manohar:1996cq}
A.~V. Manohar, \emph{{Effective field theories}},
  \href{http://dx.doi.org/10.1007/BFb0104294}{\emph{Lect. Notes Phys.} {\bf
  479} (1997) 311--362}, [\href{https://arxiv.org/abs/hep-ph/9606222}{{\tt
  hep-ph/9606222}}].

\bibitem{giudice_largest_2000}
G.~F. Giudice, E.~W. Kolb and A.~Riotto, \emph{{Largest temperature of the
  radiation era and its cosmological implications}},
  \href{http://dx.doi.org/10.1103/PhysRevD.64.023508}{\emph{Phys. Rev.} {\bf
  D64} (2001) 023508}, [\href{https://arxiv.org/abs/hep-ph/0005123}{{\tt
  hep-ph/0005123}}].

\bibitem{mazumdar_quantifying_2014}
A.~Mazumdar and B.~Zaldivar, \emph{{Quantifying the reheating temperature of
  the universe}},
  \href{http://dx.doi.org/10.1016/j.nuclphysb.2014.07.001}{\emph{Nucl. Phys.}
  {\bf B886} (2014) 312--327}, [\href{https://arxiv.org/abs/1310.5143}{{\tt
  1310.5143}}].

\bibitem{hahn_cuba_2005}
T.~Hahn, \emph{{CUBA: A Library for multidimensional numerical integration}},
  \href{http://dx.doi.org/10.1016/j.cpc.2005.01.010}{\emph{Comput. Phys.
  Commun.} {\bf 168} (2005) 78--95},
  [\href{https://arxiv.org/abs/hep-ph/0404043}{{\tt hep-ph/0404043}}].

\bibitem{Jaeckel:2016jlh}
J.~Jaeckel, V.~V. Khoze and M.~Spannowsky, \emph{{Hearing the signal of dark
  sectors with gravitational wave detectors}},
  \href{http://dx.doi.org/10.1103/PhysRevD.94.103519}{\emph{Phys. Rev.} {\bf
  D94} (2016) 103519}, [\href{https://arxiv.org/abs/1602.03901}{{\tt
  1602.03901}}].

\bibitem{Dev:2016feu}
P.~S.~B. Dev and A.~Mazumdar, \emph{{Probing the Scale of New Physics by
  Advanced LIGO/VIRGO}},
  \href{http://dx.doi.org/10.1103/PhysRevD.93.104001}{\emph{Phys. Rev.} {\bf
  D93} (2016) 104001}, [\href{https://arxiv.org/abs/1602.04203}{{\tt
  1602.04203}}].

\bibitem{Hasegawa:2019amx}
T.~Hasegawa, N.~Okada and O.~Seto, \emph{{Gravitational waves from the minimal
  gauged $U(1)_{B-L}$ model}},  \href{https://arxiv.org/abs/1904.03020}{{\tt
  1904.03020}}.

\bibitem{freeze_out_kin_mix}
A.~Banerjee, G.~Bhattacharyya, D.~Chowdhury and Y.~Mambrini{\emph{,~in
  preparation} }.

\end{thebibliography}\endgroup


\end{document}